\documentclass[letterpaper]{article} 
\usepackage{aaai24}  
\usepackage{times}  
\usepackage{helvet}  
\usepackage{courier}  
\usepackage[hyphens]{url}  
\usepackage{graphicx} 
\usepackage{natbib}  
\usepackage{caption} 
\usepackage{algorithm}
\usepackage{algorithmic}

\usepackage{newfloat}
\usepackage{listings}
\DeclareCaptionStyle{ruled}{labelfont=normalfont,labelsep=colon,strut=off} 
\floatstyle{ruled}
\newfloat{listing}{tb}{lst}{}
\floatname{listing}{Listing}
\title{GEAR-Up: Generative AI and External Knowledge-based Retrieval\\ Upgrading Scholarly Article Searches for Systematic Reviews}

\author{
    Kaushik Roy\equalcontrib\textsuperscript{\rm 1},
    Vedant Khandelwal\equalcontrib\textsuperscript{\rm 1},
    Harshul Surana\textsuperscript{\rm 1}, 
    Valerie Vera\textsuperscript{\rm 2}, 
    Amit Sheth\textsuperscript{\rm 1}, 
    Heather Heckman\textsuperscript{\rm 2}
}
\affiliations{
    \textsuperscript{\rm 1}Artificial Intelligence Institute, University of South Carolina\\
    \textsuperscript{\rm 2}Library Sciences, University of South Carolina\\

}

\usepackage{bibentry}

\begin{document}

\maketitle

\begin{abstract}
Systematic reviews (SRs) - the librarian-assisted literature survey of scholarly articles takes time and requires significant human resources. Given the ever-increasing volume of published studies, applying existing computing and informatics technology can decrease this time and resource burden. Due to the revolutionary advances in (1) Generative AI such as ChatGPT, and (2) External knowledge-augmented information extraction efforts such as Retrieval-Augmented Generation, In this work, we explore the use of techniques from (1) and (2) for SR. We demonstrate a system that takes user queries, performs query expansion to obtain enriched context (includes additional terms and definitions by querying language models and knowledge graphs), and uses this context to search for articles on scholarly databases to retrieve articles. We perform qualitative evaluations of our system through comparison against sentinel (ground truth) articles provided by an in-house librarian. The demo can be found at: https://youtu.be/zMdP56GJ9mU.
\end{abstract}

\section{Introduction}\label{sec:intro}

The proliferation of artificial intelligence (AI) technologies, e.g., Microsoft’s recent integration of ChatGPT into its Bing search architecture, showcases AI’s immense potential for shaping information search and discovery for educational purposes. For example, researchers at higher education institutions often need to review various subjects and topics systematically. For this, the researchers consult expert librarians trained in finding and evaluating the information in university libraries. AI systems have the potential to assist librarians through the systematic review process. Moreover, if developed responsibly and with input from librarians, such systems could help alleviate significant employee availability concerns (e.g., time and bandwidth limitations) \cite{borah2017analysis,bullers2018takes}. We demonstrate a pipeline to assist librarians with structured, systematic review search processes. Our method is modular and can be described as follows:

(1) \textit{Query Expansion Module} - processes queries in natural language and attaches additional context by querying external knowledge graphs and pretrained language models \cite{kawintiranon2021knowledge}. 

(2) \textit{Additional Related Query Generation Module} - Our system then uses this expanded query to prompt ChatGPT, generating queries related to the original input query \cite{lewis2020retrieval}. 

(3) \textit{Article Search and Retrieval Module} With this list of queries, we first obtain a list of articles using PubMed searches and implement a FAISS-powered retriever that narrows the search down to the most relevant articles (Title, abstracts, and relevant passages) \cite{komeili2021internet}.
\begin{figure*}[!htb]
    \centering
    \includegraphics[width=\linewidth,trim = 3cm 4.5cm 0cm 0cm, clip]{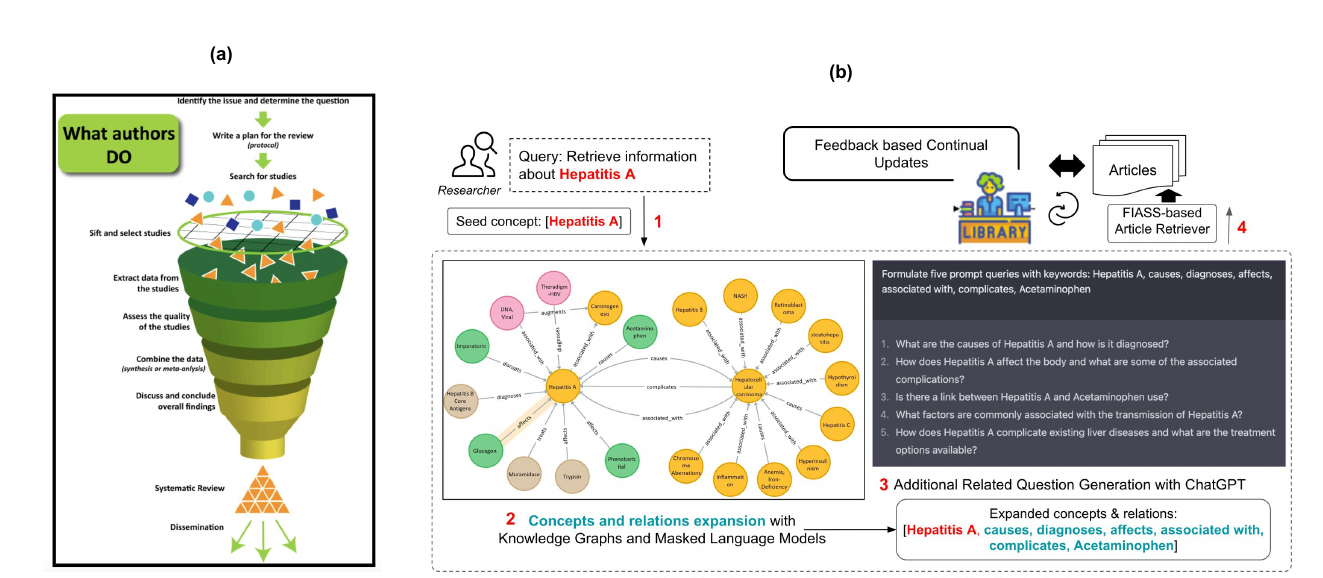}
    \caption{ \textbf{(a)} shows all the steps in the systematic review process. First, the research question is formulated in the "Identify the issue and determine the question" phase. Second, the review protocol is determined, i.e., what are the related articles to include based on query concepts and their relation to other concepts. This protocol is used to obtain a targeted search query. Lastly, the targeted search query is executed, the relevant data are extracted, the quality is assessed, and a systematic review is compiled and disseminated. Our proposed system seeks to automate the first two steps of issue identification and review protocol-based targeted query formulation. \textbf{(b)} (1) We use natural language processing tools to obtain seed concepts from the input query. In the figure example, the seed concept is Hepatitis A. (2) We query KGs (e.g., PubMed) using the seed concepts in the input query to obtain additional context. The figure shows subgraphs of concepts connected to the seed concept Hepatitis A and the relationships that connect them. We also query pretrained masked language models to obtain additional context terms. The figure shows the relationships and concepts obtained for the seed concept of Hepatitis A as causes, diagnoses, affects, associated with, complicates, and Acetaminophen. Steps (1) and (2) form the \textit{Query Expansion Module} from Section \ref{sec:intro}. (3) The concepts and relationships from the previous steps are fed into ChatGPT with an appropriate prompt to obtain a set of reformulated queries. For example, the prompt in the figure is "Formulate five prompt queries with the keywords: causes, diagnoses, affects, associated with, complicates, Acetaminophen". The figure also shows the reformulated queries that ChatGPT generates, e.g., “What are the causes of Hepatitis A and how is it diagnosed?”  Step (3) forms the \textit{Additional Related Query Generation Module} in Section \ref{sec:intro}. In Step (4), With this list of queries, we first obtain a list of articles using PubMed searches and implement a FAISS-powered retriever that narrows the search to the most relevant articles, including the article titles, abstracts, and relevant passages. This forms the \textit{Article Search and Retrieval Module} in Section \ref{sec:intro}. Finally, the list of retrieved outputs is presented to the librarian for feedback-based refinements to the overall system.}
    \label{fig:overview}
\end{figure*}
\section{The Systematic Review Process}\label{sec:srp}
The review process involves the following steps \cite{koffel2015use}: 

\textit{(i) Defining the research question}: Formulating a clear, well-defined research question of appropriate scope. Often, a researcher needs help to define a research problem precisely and interacts with the expert librarian for help with this effort. \textit{(ii) Developing a review protocol/criteria:} This step is often carried out in parallel with the first step and results in defining the terminology and topics that inform the development of the research question. \textit{ (iii) Developing inclusion and exclusion criteria:} The student needs to understand and determine whether the review will include a particular study. For this, they provide well-defined inclusion-exclusion criteria. 

Steps \textit{(i), (ii), and (iii)} correspond to; identifying the issue, determining the question, and writing a plan for the review (protocol) in Figure \ref{fig:overview} \textbf{(a)}. This functionality is carried out by modules (1) and (2) introduced in Section \ref{sec:intro}. The remaining steps involve searching a database and using existing machine learning tools to help with the later stages, including article screening, data extraction, and the risk of bias assessment. This is carried out by module (3) in Section \ref{sec:intro}. 

\section{Details of Our Implementation and Evaluation}
Due to space concerns, we ask the reader to refer to the caption of Figure \ref{fig:overview} \textbf{(b)} for details and illustration of our system and its submodules (introduced in Section \ref{sec:intro})
\paragraph{What does the Librarian Review?} The librarians' review can be incorporated at several points in the system's execution. The librarian's suggestions can be used to refine the \textit{Additional Related Query Generation  Module} to extract more contextual and relevant responses, without which ChatGPT might generate irrelevant and incoherent queries. The librarian may also analyze the \textit{safety} of the generated queries. For example, the external knowledge contains information about how acetaminophen could be misused. The system may then incorporate controls (via KG paths) to avoid showing information due to sensitive content. Similar controls can be placed to extend safety, including the relevant ethics and bias issues. Note that safety is context-sensitive. For example, it may be appropriate for addiction researchers to learn how a drug is abused (e.g., through higher doses, snorting, etc.). These aspects can be incorporated in the \textit{Query Expansion Module}.

\section{Conclusion and Future Work}
We demonstrate a system for assisting librarians in the SR process. It inputs natural language queries and leverages external knowledge to return a set of relevant research articles relevant to the review. We evaluated the retrieved articles with an in-house librarian. Our system shows favorable reviews for reducing the librarian burden by providing high-quality articles similar to a human librarian. Our work currently stands at the implementation of the three modules introduced in Section \ref{sec:intro}. Future steps involve exploring external knowledge from unstructured (e.g., documents), semi-structured (e.g., web URLs), and structured (e.g., knowledge graphs) before invoking the generative model. Furthermore, we are working on tracing back outputs to the corresponding knowledge sources to provide evidence for the appropriate parts of the system output.

\section*{Acknowledgements}
This research was built upon prior work in \cite{roy2021knowledge,roy2022tutorial,roy2022tdlr,gupta2022learning,roy2023proknow}, supported in part by NSF Award 2335967 "EAGER: Knowledge-guided neurosymbolic AI" with guardrails for safe virtual health assistants". Opinions are those of the authors and do not reflect the opinions of the sponsor \cite{sheth2021knowledge,sheth2022process,sheth2023neurosymbolic}.
\bibliography{references}
\appendix
\section{Results, Librarian Review and Evaluation}
 An in-house librarian evaluates the retrieved articles compared to \textit{sentinel}, i.e., ground-truth articles from the librarian's experience with similar queries. Figure \ref{fig:review} shows an example review. More results are provided in the supplementary.
\begin{figure}[!htb]
    \centering
    \includegraphics[trim = 0cm 2.5cm 0cm 0cm, clip, width=\linewidth]{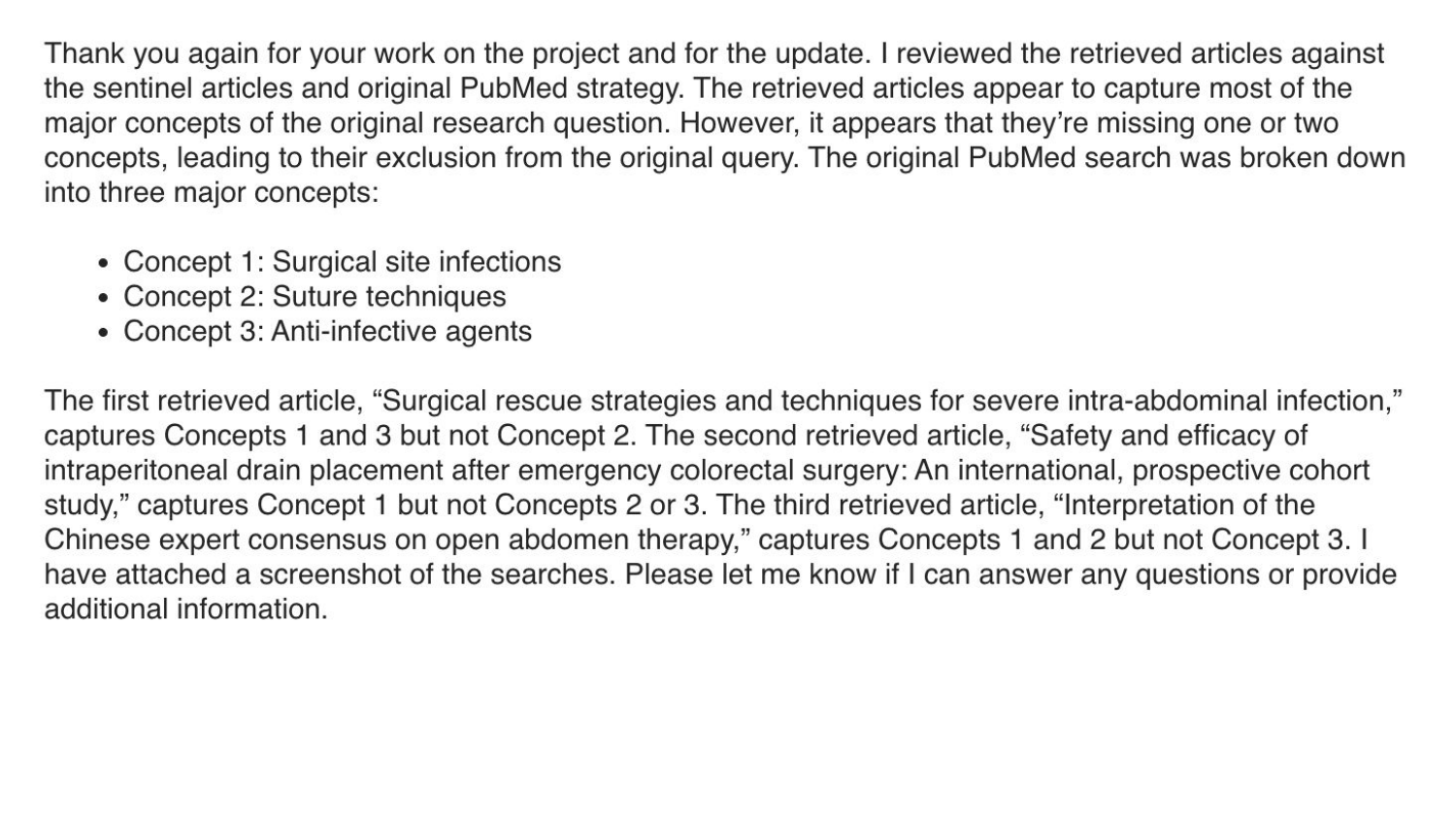}
    \caption{Example of the in-house librarian review that provides detailed comments about positives and negatives about the retrieved articles for a given query.}
    \label{fig:review}
\end{figure}

\end{document}


\appendix
\section*{Appendix A: Expansion of Mask Tokens Using MeSH Hierarchy in PubMed Searches}

\subsection*{Methodology}
The methodology involves expanding mask tokens for PubMed searches using the Medical Subject Headings (MeSH) hierarchy. This approach enhances the precision and comprehensiveness of literature searches in medical research.

\subsection*{Example Queries and MeSH Term Expansion}
A key example is the query on "Antimicrobial agents and suture techniques in preventing surgical site infections." This query involves terms like 'suture,' 'infections,' and 'prevention,' each expanded through MeSH terms to encompass a broader range of relevant literature.

\subsubsection*{Term: Suture}
Definition: Materials used in closing a surgical or traumatic wound. Similar MeSH terms include various surgical techniques and materials, illustrating the depth of the MeSH hierarchy in capturing related concepts.

\subsubsection*{Term: Infections}
Definition: Invasion of the host organism by microorganisms causing diseases. The MeSH terms expand to include various types of infections and pathogens, relevant in surgical contexts.

\subsection*{Significance of MeSH Terms in Medical Research}
MeSH terms play a crucial role in medical research by enabling comprehensive and systematic literature searches. They ensure that researchers can access a wide range of related studies, enhancing the depth and quality of systematic reviews.

\subsection*{Conclusion}
The use of MeSH terms for expanding PubMed search queries significantly contributes to the accuracy and comprehensiveness of research in medical fields. This methodology aligns with the aims of the main paper to enhance information retrieval in academic research.